\documentclass[sts,preprint]{imsart}
\RequirePackage{amsthm,amsmath,amsfonts,amssymb}
\RequirePackage[authoryear]{natbib}
\RequirePackage[colorlinks,citecolor=blue,urlcolor=blue,linkcolor=blue]{hyperref}
\RequirePackage{booktabs,array}
\startlocaldefs
\newcommand{\E}{\mathbb E}
\newcommand{\Var}{\operatorname{Var}}
\newcommand{\Cov}{\operatorname{Cov}}
\newcommand{\ind}{\mathbf 1}
\newcommand{\one}{\mathbf 1}
\newcommand{\ci}{\mathrel{\perp\!\!\!\perp}}

\newcommand{\structure}{\par\smallskip\noindent\textit{Structure.}\ }
\theoremstyle{definition}
\newtheorem{example}{Example}
\endlocaldefs
\begin{document}
\hypersetup{pdfsubject={Preprint},pdftitle={Identification multiplicity across study designs},pdfauthor={Carlos Garc\'ia Meixide and David R\'ios Insua}}
\begin{frontmatter}
\title{Identification multiplicity across study designs}
\runtitle{Identification Multiplicity}
\begin{aug}
\author[A,B]{\fnms{Carlos}~\snm{Garc\'ia Meixide}}
\author[A]{\fnms{David}~\snm{R\'ios Insua}}
\address[A]{Instituto de Ciencias Matem\'aticas (ICMAT), Spanish National Research Council (CSIC), Madrid.}
\address[B]{\newline CGM is also affiliated with the Departamento de Matem\'aticas, Universidad Aut\'onoma de Madrid, Spain.}
\end{aug}
\begin{abstract}
The answer to a causal question often admits several translations in terms of observed data. These may differ in adjustment sets, comparison groups, instruments, mediators, cutoffs, eligibility occasions, or data sources. We describe the resulting collection of analyses by three structures: the causal target that each analysis identifies under its assumptions, the assumptions that the analyses share, and the sampling dependence among their estimates. A single decomposition separates identification discrepancies from differences in causal targets, and we make both terms explicit in fourteen examples spanning adjustment, comparative panels, proxies, mediators, instruments, discontinuities, staggered adoption, sequential trials, and source combination. Comparisons within a collection can reveal differential violations of assumptions but not a bias common to all analyses; agreement is therefore evidence about the assumptions that differ, not about those that are shared. Averaging retains a causal interpretation in three ways: efficient combination of analyses that identify one target, a projection of a causal curve with a declared approximating family and weighting measure, and inference under weaker premises such as validity of at least one analysis or a known sign of its bias. Precision weights, regression-implied weights, and hierarchical models do not supply these interpretations by themselves. We relate this account to triangulation, overidentification, marginal treatment effects, transportability, and recent methods for inference under model uncertainty, and distinguish identification multiplicity from multiple testing.
\end{abstract}
\end{frontmatter}

\section{Introduction}
A study rarely supplies only one way to address a causal question. An investigator may have two plausible assumption sets that justify different covariate adjustments, several untreated populations, more than one instrument, or both experimental and observational information. Reporting only one analysis conceals these choices. Reporting every analysis makes them visible, but leaves a harder question: what have we learned from seeing them together?

Consider a clinical study of a treatment's effect on an outcome at a fixed follow-up time. One analysis adjusts for baseline severity and medical history. Another also adjusts for measurements thought to be proxies for severity. These two could be aimed at the same average treatment effect. A third compares changes in outcomes with those in an untreated population; it targets the effect among those who were treated, which coincides with the average effect only under further conditions. A fourth analysis uses variation in physicians' prescribing preferences and may instead identify an effect among patients whose treatment changes with the physician. A fifth compares treated and untreated patients among those surviving to the follow-up visit; because survival can itself depend on treatment, this comparison need not be a causal effect in any fixed population. These five estimates need not answer the same question, even if their numerical units and treatment labels agree.

There are consequently two tasks before any synthesis. We must establish what each analysis estimates under its assumptions, and decide whether those quantities coincide with the scientific target. Neither task can be delegated to a forest plot. An unusually precise estimate may concern an unusually narrow population. A striking disagreement may be an expected consequence of effect modification. Conversely, a reassuring consensus may persist because every analysis shares the same omission; as \citet{senn2024testing} remarks in an informal essay, several methods can agree and still be wrong.

We therefore study collections of analyses, each described by an assumption set, the observed data functional it uses, and the causal quantity that functional identifies when the assumptions hold. When several distinct functionals identify a common target, we speak of \emph{identification multiplicity}; when they identify different targets, of \emph{target heterogeneity}. Most collections exhibit both, and the first task of a synthesis is to tell them apart. A third situation is easily overlooked: an analysis whose causal target is not defined until the question is refined, as with the comparison among survivors (Section~\ref{sec:principal}). The word multiplicity is deliberate, but it does not make this a synonym for multiple testing. The latter concerns the calibration of a rule for making statistical claims. Identification multiplicity concerns the relation between a causal question and the observed distribution. The two meet when investigators search among analyses, select favorable results, or interpret an ensemble of tests as corroboration. Keeping them separate helps explain why neither a multiplicity correction nor a hierarchical model, by itself, settles the causal interpretation.

Several literatures address parts of this problem. Causal diagrams connect assumptions to identifying formulas \citep{pearl1995causal}. Triangulation seeks approaches with different and largely unrelated sources of bias that address the same causal question \citep{lawlor2016}. Overidentification relates restrictions on the observed distribution to specification tests and efficiency gains \citep{chensantos2018}, and evidence factors combine tests that are valid under different assumptions about treatment assignment \citep{rosenbaum2011,karmakar2019}. In econometrics, the marginal treatment effect expresses many instrument-based parameters as weighted averages of one causal curve \citep{heckmanvytlacil1999,heckman2005}, and recent work on difference-in-differences exposes the weights implied by common regressions \citep{goodmanbacon2021,dechaisemartin2020,roth2023}. Reviews of transportability and data fusion treat the alignment of sources \citep{pearlbareinboim2014,bareinboim2016,colnet2024,degtiar2023}. Multiverse and specification curve analyses display many specifications side by side \citep{steegen2016,simonsohn2020}, and recent methods provide inference when only some candidate models are valid \citep{kang2022two,yang2024robust,ghosh2025which,bhattacharya2026}.

These literatures are largely organized by design or by method. We organize them by two questions that apply to every design: what changes between the analyses in a collection, and what each operation on the collection---comparison, combination, aggregation, bounding, and testing---can establish about a causal question. Table~\ref{tab:literature} summarizes the question each tradition answers, what it takes as given, and where it enters our account.
\begin{table*}[t]
\caption{Related traditions, the question each answers, and where it enters this review. The entries describe the setting emphasized in the cited work, not every extension of it.}\label{tab:literature}
\small
\begin{tabular}{@{}>{\raggedright\arraybackslash}p{.28\textwidth}>{\raggedright\arraybackslash}p{.25\textwidth}>{\raggedright\arraybackslash}p{.27\textwidth}>{\raggedright\arraybackslash}p{.12\textwidth}@{}}
\toprule
Tradition & Question & Takes as given & Used in\\
\midrule
Triangulation \citep{lawlor2016} & Do approaches with unrelated biases point to the same conclusion? & A common causal question; known or plausible directions of bias & Sections~\ref{sec:contrasts}, \ref{sec:partial}\\
Overidentification \citep{chensantos2018} & Is the model refutable, and can efficiency be gained? & The joint model and its restrictions & Sections~\ref{sec:contrasts}, \ref{sec:average}\\
Evidence factors \citep{rosenbaum2011,karmakar2019} & Can tests valid under different assignment assumptions be combined? & A common null; (approximately) independent component tests & Section~\ref{sec:clinical}\\
Marginal treatment effects \citep{heckmanvytlacil1999,heckman2005} & What does each instrument identify, and how are the parameters related? & A latent index selection model & Section~\ref{sec:average}; Example~\ref{ex:iv}\\
Difference-in-differences with variation in timing \citep{callaway2021difference,goodmanbacon2021,dechaisemartin2020,roth2023} & What does a regression coefficient average, and which summaries are interpretable? & Parallel trends for each cohort and time & Section~\ref{sec:average}; Example~\ref{ex:staggered}\\
Transportability and data fusion \citep{pearlbareinboim2014,bareinboim2016,colnet2024,degtiar2023} & When can sources be combined for a target population? & A declared target population; transport and comparability conditions & Sections~\ref{sec:language}, \ref{sec:sources}\\
Multiverse and specification curves \citep{steegen2016,simonsohn2020} & How sensitive is a result to analytic choices? & A set of defensible specifications, organized by analytic choice & Section~\ref{sec:clinical}\\
Inference under partial validity \citep{kang2022two,yang2024robust,ghosh2025which,bhattacharya2026} & What is valid when only some candidate models hold? & At least one valid model, or a testable separation of valid and invalid ones & Sections~\ref{sec:partial}, \ref{sec:clinical}\\
This review & What changes between analyses, and what can each operation on a collection establish? & Analyses described by assumption set, functional, and target & Sections~\ref{sec:language}--\ref{sec:collection}\\
\bottomrule
\end{tabular}
\end{table*} Section~\ref{sec:language} introduces the notation. Section~\ref{sec:collection} develops the account: what contrasts can and cannot reveal, how sampling dependence enters, when averaging retains a causal interpretation, and what remains under partial validity. Sections~\ref{sec:same}--\ref{sec:sources} then work through fourteen designs. Each example gives the observed data, a target, sufficient identifying conditions, and the resulting functional, and closes with the structure of the collection it generates: the target of each analysis, its discrepancy term, and what the analyses share. These conditions describe an idealized design; their credibility must be defended in the application. Section~\ref{sec:clinical} turns to multiplicity in clinical evidence, and Section~\ref{sec:discussion} concludes.

\section{A common language}
\label{sec:language}
Let $Q$ denote a full data law that contains the potential outcomes needed to define the causal question, and let $P$ denote the law that $Q$ induces on the observed data $O$ under the stated sampling model. For a binary treatment $A$ and outcome $Y$, a familiar target is
\begin{equation}\label{eq:target}
 \theta(Q)=\E_Q(Y^1-Y^0).
\end{equation}
Here $Y^a$ is the outcome under treatment $a$. We suppress $Q$ when the meaning is clear. Consistency, $Y=Y^A$, refers to the treatment versions and follow-up time specified in the question. Interference is excluded in the examples unless incorporated into the intervention definition.

An analysis strategy $j$ consists of an assumption set $\mathcal A_j$, an observed data functional $\psi_j$, and a causal quantity $\theta_j(Q)$ such that
\begin{equation}\label{eq:identification}
 Q\in\mathcal A_j \quad\Longrightarrow\quad \psi_j(P)=\theta_j(Q).
\end{equation}
We call a set of such triples, $(\mathcal A_j,\psi_j,\theta_j)$ for $j=1,\ldots,J$, a collection of analyses. The estimator $\widehat\psi_j$ is a further object: a statistic intended to estimate $\psi_j(P)$. This distinction is useful even when assumptions fail. A well-defined $\psi_j(P)$ can still be estimated, but its causal interpretation need not survive. Statistical support and existence conditions remain necessary for the formula itself.

Two expressions that agree for every $P$ in their domain define one functional. For example, outcome regression and inverse probability representations of adjustment for a fixed covariate set are the same map of $P$. Different estimators of that map can disagree because of finite samples, regularization, or misspecified nuisance models. Such disagreement is worth studying, but it is not a second identification argument, and the corresponding estimates should not be counted again. Conversely, two distinct maps may agree only on distributions satisfying shared restrictions. This is precisely where comparisons can carry information about those restrictions.

For a scientific target $\theta$, the decomposition
\begin{equation}\label{eq:decomp}
 \psi_j(P)-\theta
 =\underbrace{\psi_j(P)-\theta_j}_{\text{identification discrepancy } b_j}
  +\underbrace{\theta_j-\theta}_{\text{target difference}}
\end{equation}
separates two reasons for disagreement. It is a conceptual decomposition, not an assertion that the two terms are separately observable. A target difference can reflect a change in population, treatment version, time origin, follow-up, outcome, or effect scale. The local effects in a discontinuity design illustrate the point: different cutoffs can identify different causal effects even when every continuity assumption is correct. Standardization can sometimes align heterogeneous targets; additional transport assumptions are then part of identification, rather than a cosmetic adjustment to the estimates.

\emph{Sampling models.} In most examples the data are $n$ independent, identically distributed units from $P$. Two departures recur. In designs with one treated unit or series (Example~\ref{ex:synthetic} and the interrupted time series discussed after it), the expectations below are defined only under a stated model for how the series could have varied, such as a time series or design-based model, and $P$ is the law of the observed panel under that model. In designs that combine separately collected sources (Examples~\ref{ex:transport}--\ref{ex:surrogate}), the data are independent samples from source-specific laws $P_s$, possibly with a target covariate sample; $P$ then denotes this collection together with the sampling fractions, and $Q$ must describe the populations from which the sources are drawn. Following \citet{ung2026}, we treat the target population and the sampling model as part of the question.

For most examples, $X$ denotes baseline covariates, $A$ treatment, and $Y$ outcome. A source or population label is $S$, an instrument is $Z$, and a mediator is $M$. In longitudinal settings, $t$ indexes time and $G$ denotes treatment initiation time. The index $j$ always labels an analysis strategy. Expectations without a subscript are taken under $P$; expectations involving potential outcomes refer to $Q$. We use sums for discrete variables, with integrals understood for continuous variables. All conditional means are assumed integrable on the supports where they are evaluated.

Table~\ref{tab:map} gives a map of the examples. The organizing question is what changes between analyses. This is more informative than counting the number of available estimates.
\begin{table*}[t]
\caption{What changes between analyses? A common target requires the conditions in the final column, in addition to identification within each analysis.}\label{tab:map}
\small
\begin{tabular}{@{}>{\raggedright\arraybackslash}p{.15\textwidth}>{\raggedright\arraybackslash}p{.21\textwidth}>{\raggedright\arraybackslash}p{.26\textwidth}>{\raggedright\arraybackslash}p{.30\textwidth}@{}}
\toprule
Design family & Source of multiplicity & Target of each analysis & Requirement for a common target\\
\midrule
Adjustment & Covariate sets & Population average effect & Same population, intervention, and outcome\\
Comparative panels & Controls or counterfactual trend models & Effect on the treated group or unit & Same treated population and untreated intervention\\
Proxies; front-door & Variable roles or mediator sets & Population average effect & Every route identifies the same total effect\\
Instruments & Encouragements or instruments & Complier effect for each instrument & Alignment of local effects or justified homogeneity\\
Discontinuities & Sites, cutoffs, or kinks & Effect at each cutoff and site & Alignment of local populations and margins\\
Staggered adoption & Cohorts and follow-up times & Effect for cohort $g$ at time $t$ & Declared aggregation; homogeneity is additional\\
Sequential trials & Eligibility occasions & Effect among those eligible at $t$ & A specified distribution of people or person-occasions\\
Multiple sources & External controls or trials & Effect in a declared target population & Standardization and cross-source comparability\\
Surrogate designs & Surrogate sets or validation sources & Long-term effect in the trial population & Valid surrogate bridges for a common long-term outcome\\
\bottomrule
\end{tabular}
\end{table*}

\section{What a collection of analyses can establish}
\label{sec:collection}
A collection has at least three structures: the map from analyses to targets, the overlap in identifying assumptions, and the dependence among estimates computed from shared data. Each structure answers a different question, and each operation on the collection relies on a different one. This section states the general results; the examples in Sections~\ref{sec:same}--\ref{sec:sources} supply their instances.

\subsection{What contrasts can and cannot reveal}
\label{sec:contrasts}
For aligned targets, $\theta_j=\theta$ and $b_j=\psi_j(P)-\theta$. An observed difference estimates $b_j-b_k$, not either bias separately. More generally, with $\boldsymbol\psi=(\psi_1,\ldots,\psi_J)^\top$ and any contrast matrix $C$ satisfying $C\one=0$,
\begin{equation}\label{eq:contrasts}
 C\boldsymbol\psi=C\boldsymbol b.
\end{equation}
Adding a common number $c$ to every $b_j$ leaves this contrast unchanged. Equality of all functionals therefore cannot identify their distance from the causal target. This is not a failure of statistical power. It is information that these comparisons do not contain. With several comparison groups (Example~\ref{ex:controls}), for instance, contrasts reveal differences between untreated trends, not a departure shared by every group.

The clinical meaning is straightforward. Three analyses that omit an unmeasured severity variable can agree. Three external sources can share an obsolete standard of care (Example~\ref{ex:external}). Three mediator sets can all miss the same direct path (Example~\ref{ex:frontdoor}). An independent measurement, a credible experimental anchor, or a restriction on the common component of bias (Section~\ref{sec:partial}) supplies information of a different kind. More variants of the same assumption do not.

If all strategies identify one target, their simultaneous validity implies equality. A well-calibrated rejection of equality can therefore challenge that joint claim. It does not name the invalid strategy. If the targets differ, rejection can instead be expected under valid identification, as with instruments that move different people (Example~\ref{ex:iv}) or sites with different cutoffs (Example~\ref{ex:rd}). And equality is only a necessary consequence of the joint causal model, not a sufficient characterization of it. Overidentification theory formalizes connections between observable restrictions, testing, and efficiency \citep{chensantos2018}; it does not turn passing a specification test into proof of causal validity.

A helpful design exercise is to record which assumptions each analysis shares. Two adjustment sets may share one poorly measured confounder. Several instrumental variable ratios may rely on the same exclusion argument. A front-door and an adjustment analysis may have different confounding assumptions while sharing the same outcome misclassification. Such a record makes the triangulation principle of \citet{lawlor2016} operational: what matters is diversity in plausible failure mechanisms, not diversity in method names. Joint assertions also need scrutiny: a set of jointly asserted graphs may impose restrictions far stronger than any one graph, and may even be scientifically incoherent.

\subsection{Sampling dependence}
\label{sec:dependence}
When $O_1,\ldots,O_n$ are independent, identically distributed sampling units and $J$ is fixed, suppose each estimator admits the joint expansion
\begin{equation}\label{eq:if}
 \widehat{\boldsymbol\psi}-\boldsymbol\psi
 =\frac1n\sum_{i=1}^n\boldsymbol\phi(O_i)+o_p(n^{-1/2}),
\end{equation}
with $\E\boldsymbol\phi=0$ and finite covariance $\Sigma=\E(\boldsymbol\phi\boldsymbol\phi^\top)$.
The covariance of the estimator vector is then approximately $V=\Sigma/n$. Joint influence functions or resampling of whole independent units can estimate it. \citet{kennedy2023} develops the estimation perspective behind such expansions. For repeated trial eligibility (Example~\ref{ex:trials}), the independent unit is typically a person, not a person-trial record. Site-level assignment or common shocks (Example~\ref{ex:bartik}) may require a different asymptotic framework altogether, as may the single series designs of Example~\ref{ex:synthetic}. Boundary estimation, weak instruments, and unstable inverse problems also need their own treatment; \eqref{eq:if} is not asserted for every design in this review.

Dependence changes the interpretation of visual disagreement. For two estimates,
\[
 \Var(\widehat\psi_j-\widehat\psi_k)
 =V_{jj}+V_{kk}-2V_{jk}.
\]
Analyses that share the treated units, such as difference-in-differences with several comparison groups, typically have large positive covariances. A small difference between highly correlated estimates can be precisely estimated even when both marginal intervals are wide. Conversely, the number of overlapping marginal intervals is not a calibrated measure of consistency.

For a fixed full row rank contrast matrix $C$ and nonsingular $C\Sigma C^\top$, a joint Wald test of $C\boldsymbol\psi=0$ uses
\[
 n(C\widehat{\boldsymbol\psi})^\top
 (C\widehat\Sigma C^\top)^{-1}
 (C\widehat{\boldsymbol\psi})
 \ \Longrightarrow\ \chi^2_{\operatorname{rank}(C)}.
\]
This statement requires the joint central limit theorem and consistent covariance estimation. Duplicated or asymptotically equivalent analyses can make the contrast covariance singular. Removing redundant contrasts is preferable to counting them as additional confirmation.

\subsection{Combining, aggregating, and projecting}
\label{sec:average}
Averages of analyses can be justified in three different ways, and the justification determines what the average means.

\emph{Combination under a common target.} If every functional identifies the same $\theta$, any fixed weights $w$ with $\one^\top w=1$ preserve that target. For nonsingular $V$, the minimum variance linear weights are
\begin{equation}\label{eq:gls}
 w_{\mathrm{GLS}}=\frac{V^{-1}\one}{\one^\top V^{-1}\one}.
\end{equation}
This is an efficiency calculation under exact alignment and validity. Negative weights are possible; nonnegative constraints give another optimization problem. Estimated weights require their own stability and inferential justification. Neither the formula nor its precision estimates the probability that a strategy is causally valid.

\emph{Projection of a causal curve.} When analyses identify different causal quantities, an aggregate has a causal meaning only if it summarizes those quantities in a declared way. Let $v$ range over an index set $\mathcal V$ and let $v\mapsto\theta(v)$ be a causal curve: each $\theta(v)$ is a causal parameter of $Q$. A projection specifies an approximating family $\{m_\beta:\beta\in B\}$ and a weighting measure $\nu$ on $\mathcal V$, and targets
\begin{equation}\label{eq:projection}
\begin{aligned}
 \beta^\ast&=\arg\min_{\beta\in B}\int r_\beta(v)^2\,d\nu(v),\\
 r_\beta(v)&=\theta(v)-m_\beta(v),
\end{aligned}
\end{equation}
assuming $\int\theta(v)^2\,d\nu(v)<\infty$ and a unique minimizer. For the constant family, $\beta^\ast=\int\theta\,d\nu$. If $B_1,\ldots,B_J$ partition the target population, $\theta_j=\E(Y^1-Y^0\mid B_j)$, and $\nu$ puts mass $P(B_j)$ on $j$, this gives
\begin{equation}\label{eq:partition}
 \theta=\sum_j P(B_j)\theta_j.
\end{equation}
The scientific meaning of $\beta^\ast$ comes from the curve and from $\nu$, not from the existence of the least squares representation. Three consequences follow. First, the curve must be identified on the support of $\nu$; a projection does not extrapolate where no analysis identifies $\theta(v)$. Second, unless the family is saturated, $\beta^\ast$ depends on $\nu$, so a misspecified family summarizes the curve differently under different measures. Third, the index $v$ must be stated. It may index populations, such as complier groups, cutoffs, cohorts, or eligible occasions; interventions, such as dose, timing, or time since adoption; or both, as in staggered adoption, where the cohort $g$ indexes a population and $t-g$ the duration of exposure. Alternative identification routes for the same $\theta(v)$ are not values of $v$: averaging them is combination under a common target, not projection.

Several familiar summaries are projections of this kind. Under a latent index selection model, with $u$ indexing resistance to treatment and $\mathrm{MTE}(u)$ the marginal treatment effect, the average effect, the effect on the treated at propensity $p$, and the complier effect for a shift from $p'$ to $p$ are
\[
\begin{gathered}
 \int_0^1\mathrm{MTE}(u)\,du,\qquad
 \frac1p\int_0^p\mathrm{MTE}(u)\,du,\\
 \frac{1}{p-p'}\int_{p'}^{p}\mathrm{MTE}(u)\,du,
\end{gathered}
\]
that is, constant projections of one curve under three measures \citep{heckmanvytlacil1999,heckman2005}. Group time effects in staggered adoption can be summarized by a policy distribution over cohorts and times, or projected onto an event time family $m_\beta(g,t)=\beta_{t-g}$ (Example~\ref{ex:staggered}). The two eligibility summaries of Example~\ref{ex:trials} are constant projections under a uniform measure over occasions and under the distribution of eligible person-occasions.

Many common estimators instead carry a measure that no one chose. Precision weights put mass proportional to the entries of $V^{-1}\one$. A two-way fixed effects coefficient is a weighted sum of group time effects whose weights can be negative \citep{dechaisemartin2020,goodmanbacon2021}. Pooling sites with different cutoffs after centering the running variable weights each cutoff's effect by the density of the running variable at that cutoff times the site's probability \citep{cattaneo2016multiple}. A pooled partially linear regression over sequential trials weights trial-specific effects by the conditional variance of treatment \citep{gervasoni2026}. The weights of a shift-share regression can be negative (Example~\ref{ex:bartik}). Each of these is a well-defined functional of $P$; none is a causal summary unless its implied measure has been justified.

A small constructed example makes the distinction concrete. Suppose two equally prevalent patient groups have effects $0$ and $2$, so the population average effect is $1$. Independent unbiased subgroup estimates have variances $0.01$ and $1$. Inverse variance weights are approximately $0.9901$ and $0.0099$, giving an expected pooled value of $0.0198$. The resulting estimator is precise for that precision-weighted effect, but it does not estimate the population average. Nothing has gone wrong with either subgroup analysis. The error is the interpretation of the weights.

The effect scale matters too. With subgroup risks $r_{aj}=P(Y^a=1\mid B_j)$ and population weights $q_j$, the marginal risk difference is $\sum_jq_j(r_{1j}-r_{0j})$. The marginal odds ratio is
\[
 \frac{\bar r_1/(1-\bar r_1)}{\bar r_0/(1-\bar r_0)},
 \qquad \bar r_a=\sum_jq_jr_{aj},
\]
not a prevalence-weighted average of subgroup odds ratios. One must aggregate the appropriate potential outcome distributions before applying a nonlinear contrast. Changes in conditional and marginal odds ratios need not diagnose confounding.

\emph{Partial pooling.} A third reason to average is to estimate a vector of related effects by sharing information. A hierarchical model can express similarities across real populations, trials, or sites. \citet{gelman2012} explain how partial pooling changes the estimation problem in multiple comparison settings by modeling relations among effects. If the target is a population average, posterior or shrunken cell estimates can be combined with the population weights. A model for a collection of identification discrepancies has a different interpretation: its center is not automatically the causal effect, and a hand-built catalog is not a random sample of strategies merely because it is large.

The most direct way to see the issue is, for weights summing to one,
\[
 \sum_jw_j\psi_j-\theta
 =\sum_jw_j(\theta_j-\theta)+\sum_jw_jb_j.
\]
Target alignment, or a declared projection, controls the first term. Causal validity or an independently justified bias restriction controls the second. Variance minimization controls neither. In particular, a condition that biases average to zero must come from scientific or design information. It is not implied by small dispersion, exchangeable labels, or an empirical Bayes fit.

\subsection{If only some strategies are credible}
\label{sec:partial}
There is a consequential difference between asserting all $\mathcal A_j$ and asserting that at least one holds. Under the latter claim, the causal target belongs to the candidate set $\{\psi_1(P),\ldots,\psi_J(P)\}$ for aligned strategies. Its average is not generally the target. If $I_j$ has coverage at least $1-\alpha$ whenever strategy $j$ is valid, the union $\bigcup_j I_j$ covers $\theta$ with at least that probability whenever at least one strategy is valid: the union contains the interval supplied by a valid strategy. This construction can be wide or disconnected, but it honestly reflects the weaker premise. It requires a fixed candidate collection and the stated validity of each interval under its own model. \citet{kang2022two} use unions of this kind for instruments when the identity of the valid instruments is unknown. For candidate adjustment sets, \citet{ghosh2025which} show that unions and convex hulls of intervals can have width that does not shrink with the sample size; their remedy shifts the target to a reweighted population on which the candidate sets are compatible. That is a deliberate change of estimand, and it should be reported as one.

Bias restrictions can refine the analysis. If $|b_j|\le B_j$ is scientifically defensible for every strategy, then population values imply
$\theta\in\bigcap_j[\psi_j-B_j,\psi_j+B_j]$; sampling uncertainty must also be incorporated. An empty intersection challenges the restrictions jointly. If bounds are defensible for only one unknown strategy, a union is appropriate instead. Restrictions on the sign of a discrepancy are often easier to defend than restrictions on its size. For aligned strategies with $b_1\ge0\ge b_2$, the population values give $\theta\in[\psi_2,\psi_1]$. This is the formal content of the recommendation of \citet{lawlor2016} to seek approaches whose biases are expected to point in opposite directions. \citet{dingli2019} establish such a bracketing relation for the difference-in-differences and lagged outcome analyses of Example~\ref{ex:panel}: under the conditions stated there, each analysis errs in a known direction when the other's assumption holds. A sign restriction addresses the common component of bias that comparisons alone cannot see.

Methods that assume a majority or a plurality of valid instruments, or structured pleiotropy, introduce still other premises (Example~\ref{ex:iv}). These are different inferential problems, not interchangeable ways of averaging an inconvenient forest plot. \citet{bhattacharya2026} develop weights based on testable implications of candidate models and bound the resulting causal discrepancy. Their framework requires conditions under which diagnostics separate valid from invalid models. An invalid model that passes the available diagnostics remains a substantive difficulty; a diagnostic weight is not an assumption-free probability of validity.

\section{Several routes to the same effect}
\label{sec:same}
We first hold the population and intervention fixed. The alternatives differ in what they assume about the missing counterfactual, not in whose effect they seek.

\subsection{Adjustment sets}
\begin{example}[Several sufficient covariate sets]\label{ex:adjustment}
Observe $O=(L,A,Y)$, where $L$ is a vector of baseline measurements. For each candidate subvector $X_j\subseteq L$, define
\begin{equation}\label{eq:adjust}
\begin{aligned}
 \psi_j(P)&=\E\{\mu_{1j}(X_j)-\mu_{0j}(X_j)\},\\
 \mu_{aj}(x)&=\E(Y\mid A=a,X_j=x).
\end{aligned}
\end{equation}
If $Y^a\ci A\mid X_j$, consistency holds, and $0<P(A=1\mid X_j)<1$ almost surely, this functional identifies the population average treatment effect \eqref{eq:target}. Different sufficient sets need not contain the same variables \citep{pearl1995causal}.
\end{example}
The outer expectation matters. Every version of \eqref{eq:adjust} averages over the same study population, each through the marginal distribution of its own covariates. Effect modification alone does not make these adjusted effects different. By contrast, trimming each analysis to its own overlap population can change the target. A comparison that was intended to assess confounding assumptions can then become a comparison of populations.

More adjustment is not automatically better. A baseline variable can be a collider, an instrument, or a noisy proxy, and adding it can change the required assumptions or the precision. The relevant justification is its causal role, not simply its association with treatment or outcome. For a fixed $X_j$, replacing regression standardization with weighting does not create a new route: with $e_j(x)=P(A=1\mid X_j=x)$ bounded away from 0 and 1,
\[
 \psi_j(P)=\E\left\{\frac{AY}{e_j(X_j)}-
                    \frac{(1-A)Y}{1-e_j(X_j)}\right\}.
\]
This equality follows by iterated expectation, without invoking causal assumptions; it uses only the support condition on $e_j$. It separates an algebraic identity from the causal equality in \eqref{eq:identification}.
\structure All analyses target the population average effect, provided none is trimmed to its own overlap population. The discrepancy $b_j$ is the confounding left by $X_j$. Sets that omit the same confounder, or mismeasure it in the same way, share that part of their discrepancy.

\subsection{Control groups and counterfactual trends}
\begin{example}[Several untreated comparison groups]\label{ex:controls}
Observe $O=(C,Y_0,Y_1)$, with $C=0$ labeling a group treated between periods 0 and 1 and $C=j$, $j=1,\ldots,J$, labeling untreated groups. Let $\Delta Y=Y_1-Y_0$ and fix
$\theta=\E(Y_1^1-Y_1^0\mid C=0)$.
Under no anticipation, consistency, and the parallel trends restriction
\[
 \E(Y_1^0-Y_0^0\mid C=0)=\E(Y_1^0-Y_0^0\mid C=j),
\]
the functional
\begin{equation}\label{eq:did}
 \psi_j(P)=\E(\Delta Y\mid C=0)-\E(\Delta Y\mid C=j)
\end{equation}
identifies the same effect on the treated for every valid comparison group. Each group must have positive probability. A conditional version replaces the second term with the comparison group conditional trend averaged over the treated group's $X$ distribution, with corresponding overlap and conditional parallel trends \citep{callaway2021difference}.
\end{example}
The comparison groups must represent the same untreated intervention. Patients receiving another active therapy are not interchangeable with untreated patients merely because both are called controls. In a clinical application, eligibility, treatment initiation, follow-up, and concomitant care must be aligned before comparing trends. Likewise, an untreated geographical neighbor may share a policy spillover; geographic proximity is neither a guarantee of parallel trends nor a guarantee against contamination.
\structure Every analysis targets the effect on the treated group. Writing $d_j=\E(Y_1^0-Y_0^0\mid C=j)$, the discrepancy is exactly $b_j=d_0-d_j$, so $\psi_j-\psi_k=d_k-d_j$: contrasts remove the treated group's unknown counterfactual trend and measure differential trend violations, not a departure from $d_0$ shared by every comparison group. All analyses share the treated group's data, so their estimates are positively correlated.

\begin{example}[Different models for one panel]\label{ex:panel}
Keep the convention of Example~\ref{ex:controls} with two groups: $C=0$ is treated between periods 0 and 1, and $C=1$ remains untreated. The same treated group effect can be approached through parallel trends, adjustment for the baseline outcome, or a changes-in-changes model. Under no anticipation,
$\E(Y_1^0\mid Y_0,C=0)=\E(Y_1^0\mid Y_0,C=1)$, and inclusion of the treated group's baseline support in the untreated group's, baseline outcome adjustment yields
\[
\begin{aligned}
 \psi_{\mathrm{lag}}(P)&=\E\{Y_1-m_1(Y_0)\mid C=0\},\\
 m_1(y)&=\E(Y_1\mid Y_0=y,C=1).
\end{aligned}
\]
Alternatively, suppose the untreated outcome is $Y_t^0=h_t(U)$, with $h_t$ strictly increasing and common across groups, the distribution of $U$ within each group stable over time, and continuously distributed outcomes. With no anticipation and the required support inclusion, the changes-in-changes transformation is
\[
\begin{aligned}
 T(y)&=F_{11}^{-1}\{F_{10}(y)\},\\
 \psi_{\mathrm{CIC}}(P)&=\E\{Y_1-T(Y_0)\mid C=0\},
\end{aligned}
\]
where $F_{1t}$ is the untreated group's outcome distribution at time $t$ \citep{athey2006}. With discrete outcomes the same model yields bounds rather than a point value. These are alternative assumptions about the same counterfactual mean, when the target is kept fixed.
\end{example}
The models need not be nested. Parallel trends restricts average untreated changes; baseline adjustment restricts a conditional mean; changes-in-changes restricts the evolution of a latent ordering. Nor does the availability of a distributional model require changing the question: the changes-in-changes counterfactual distribution can be summarized by the mean in the display. A quantile contrast, if chosen instead, is a different estimand.
\structure One target, three non-nested restrictions, the same treated and untreated units. Agreement cannot be read as three independent replications. The bracketing relation of Section~\ref{sec:partial} adds sign information for the first two analyses that a contrast cannot supply.

\begin{example}[Alternative synthetic controls]\label{ex:synthetic}
Suppose one unit is treated after time $T_0$ and donor units remain untreated, and let expectations refer to a stated model for how the panel could have varied. For a donor set and pretreatment fitting rule indexed by $j$, let $w_{jk}(P)$ be nonnegative weights summing to one, defined from pretreatment quantities with a fixed tie-breaking rule. For a posttreatment time $t$, write $\mu_{kt}^0=\E(Y_{kt}^0)$ and define
\[
 \psi_{jt}(P)=\E(Y_{0t})-\sum_k w_{jk}(P)\E(Y_{kt}).
\]
If donors are unaffected by the intervention and
$\mu_{0t}^0=\sum_k w_{jk}(P)\mu_{kt}^0$, this functional identifies
$\theta_t=\E(Y_{0t}^1-Y_{0t}^0)$.
The condition is stated directly on the counterfactual mean; its substantive content lies in why pretreatment fit should support that posttreatment representation \citep{abadie2010}.
\end{example}
A different donor pool or fitting window can change the functional. An equivalent implementation of the same optimization does not. The displayed target is an expected unit effect, whereas \citet{abadie2010} target the realized effect $Y_{0t}^1-Y_{0t}^0$ of the treated unit; inference for the realized effect requires an appropriate design or time series formulation. Neither follows from treating donor units as independent copies of the treated unit. Good pretreatment fit is evidence about fit, not a proof of counterfactual stability.
\structure All donor sets target the same unit and time. The discrepancy is $\mu_{0t}^0-\sum_kw_{jk}\mu_{kt}^0$. Analyses share the treated series and usually many donors, and with one treated series the dependence among them is governed by the time series or design model rather than by \eqref{eq:if}.

Interrupted time series with control series fit this template without a model for the series. With a single control series $j$ and $\mu_{jt}^0=\E(Y_{jt}^0)$, the analogous restriction is that the untreated gap $\mu_{0t}^0-\mu_{jt}^0$ after $T_0$ keeps its value at a pretreatment time $s$, which gives
\[
 \psi_{jt}(P)=\E(Y_{0t})-\E(Y_{jt})-\{\E(Y_{0s})-\E(Y_{js})\},
\]
a parallel trends restriction across series, as in Example~\ref{ex:controls} with control series in place of control groups. The restriction extends a pretreatment relationship beyond the intervention date. A control must capture the concurrent events that threaten the comparison while remaining unaffected by the intervention; a second outcome in the treated population may serve this purpose only if its exclusion and relation to the relevant shocks are credible \citep{lopez2018}. Seasonality, changing measurement, and serial dependence must also be addressed. Changing the postintervention horizon changes $\theta_t$; changing a valid control for one horizon does not. Controls exposed to the same concurrent event share their discrepancy.

\subsection{Proxies and mediating paths}
\begin{example}[Alternative proxy roles]\label{ex:proximal}
Observe $O=(L,A,Y)$. A role assignment partitions baseline measurements into $(X_j,Z_j,W_j)$, where $Z_j$ is a treatment side proxy and $W_j$ an outcome side proxy for an unobserved confounder $U$. Suppose there is an integrable bridge $h_j$ satisfying
\begin{equation}\label{eq:bridge}
 \E(Y\mid Z_j,A,X_j)=\E\{h_j(W_j,A,X_j)\mid Z_j,A,X_j\}.
\end{equation}
Sufficient causal conditions include consistency, latent treatment exchangeability $Y^a\ci A\mid(U,X_j)$, latent positivity, and the proxy restrictions
\[
 Y\ci Z_j\mid(U,A,X_j),\qquad W_j\ci(Z_j,A)\mid(U,X_j).
\]
In addition, require completeness: for each $(a,x)$ and square-integrable $g$,
\[
 \E\{g(U)\mid Z_j,A=a,X_j=x\}=0
 \ \Longrightarrow\ g(U)=0
\]
almost surely. With a square-integrable bridge and the requisite support conditions, these assumptions yield
\begin{equation}\label{eq:proximal}
 \psi_j(P)=\E\{h_j(W_j,1,X_j)-h_j(W_j,0,X_j)\}=\theta
\end{equation}
\citep{miao2018identifying,cui2024semiparametric,tchetgen2024introduction}. We restrict the candidate domain to distributions where a bridge exists and its mean contrast is well defined; uniqueness of the observed bridge can be imposed separately.
\end{example}
Here the variables are fixed and their proposed roles change. A variable that affects the outcome directly may be admissible as an outcome side proxy yet inadmissible as a treatment side proxy. A measurement of previous adherence, for example, requires a temporal and clinical argument; calling it a proxy does not supply an exclusion restriction. Different role assignments are competing causal hypotheses, not a license to permute columns until an attractive answer appears.

The bridge equation is an inverse problem. Existence, causal identification of its mean, uniqueness of a solution, and stable estimation are separate questions. A large catalog of assignments does not settle any of them.
\structure Every assignment targets the population average effect. A wrongly classified variable produces a discrepancy, and assignments that share it may share the discrepancy; combinatorially many assignments do not provide combinatorially many independent sources of evidence.

\begin{example}[Alternative front-door sets]\label{ex:frontdoor}
Observe $O=(X,A,M,Y)$, and let $M_j$ be a candidate mediator or mediator vector. Conditional on $X$, suppose $M_j$ intercepts every directed path from $A$ to $Y$, there is no unblocked back-door path from $A$ to $M_j$, and $A$ blocks every back-door path from $M_j$ to $Y$. Assume consistency and positivity on the required cross-treatment supports. Writing $m_j(a,m,x)=\E(Y\mid A=a,M_j=m,X=x)$, set
\begin{equation}\label{eq:frontdoor}
\begin{aligned}
 r_{aj}(P)&=\E_X\!\int \sum_{a'=0}^1 m_j(a',m,X)\\
 &\quad\times P(A=a'\mid X)\,dP(m\mid A=a,X),\\
 \psi_j(P)&=r_{1j}(P)-r_{0j}(P).
\end{aligned}
\end{equation}
Then $r_{aj}(P)=\E(Y^a)$ and $\psi_j(P)=\theta$ \citep{pearl1995causal}.
\end{example}
Full interception is a strong condition. Two mediators on separate parallel causal paths do not each satisfy it merely because both are associated with the outcome. Alternative complete mediator sets can exist, but each requires its own causal justification. The generalized front-door criterion of \citet{fulcher2020robust} allows a direct effect and identifies the population intervention indirect effect; that is a change of target, not an automatic repair of the original claim about the total effect.

Adjustment, proxy, and front-door strategies can coexist in one dataset. Their appeal for triangulation is that they may fail for different reasons, and an analyst should distinguish claiming that every route is valid from the weaker belief that at least one is valid (Section~\ref{sec:partial}).
\structure Complete mediator sets target the total effect. A set that misses a direct path has a discrepancy that is not generally zero, and sets that miss the same path share it.

\section{When the design changes the effect}
\label{sec:local}
The next designs are often presented as generating many estimates of ``the treatment effect.'' Their identifying results usually deliver a different quantity: an effect in another population, at another margin, or under another intervention. Preserving that difference is essential to interpreting multiplicity.

\subsection{Instruments, encouragements, and local populations}
\begin{example}[Several binary instruments]\label{ex:iv}
Observe $O=(Z_1,\ldots,Z_J,A,Y)$ with binary instruments. Let $A(z_1,\ldots,z_J)$ denote treatment when the instruments are set to $(z_1,\ldots,z_J)$, and assume exclusion: the instruments affect $Y$ only through $A$, so that $Y^a$ is well defined. For instrument $j$, let $A_j(z)=A(z,Z_{-j})$ be treatment when $Z_j$ is set to $z$ and the other instruments keep their realized values. Suppose $Z_j$ is independent of $(Y^0,Y^1,A_j(0),A_j(1))$, which holds, for instance, when the instruments are jointly independent of the potential variables and $Z_j$ is independent of $Z_{-j}$. Suppose also that $Z_j$ has positive probability at both levels and that $A_j(1)\ge A_j(0)$ almost surely. If its first stage is nonzero, the Wald functional
\begin{equation}\label{eq:wald}
\begin{aligned}
 \psi_j(P)&=\frac{\E(Y\mid Z_j=1)-\E(Y\mid Z_j=0)}
                  {\E(A\mid Z_j=1)-\E(A\mid Z_j=0)}\\
 &=\E\{Y^1-Y^0\mid A_j(1)>A_j(0)\}
\end{aligned}
\end{equation}
identifies an instrument-specific complier effect \citep{imbens1994late}: the effect among units whose treatment responds to $Z_j$ given the realized values of the other instruments.
\end{example}
Monotonicity in each instrument separately is already a restriction on joint choice behavior. \citet{mtw2021} show that the usual monotonicity condition for several instruments effectively imposes homogeneous choice behavior, and give weaker partial monotonicity conditions under which two-stage least squares is a positively weighted average of complier effects. Different encouragements can move different people. Physician preference, distance to care, and a coverage rule need not identify the same effect. Even perfect random assignment of each encouragement does not align its complier population with the whole population. A common effect follows from additional restrictions, such as a constant treatment effect; it does not follow from exclusion alone.

Genetic variants in Mendelian randomization are instruments of this kind, and the same conditions apply variant by variant. A variant with a direct path to the outcome violates exclusion. A pathway shared by many variants moves their ratios together and is invisible to their contrasts. Even a valid variant need not identify the effect of a specified clinical intervention: a lifelong genetic perturbation and a treatment initiated at diagnosis can differ in duration, pathway, and population. With a continuous exposure, the ratio is not automatically a binary treatment complier effect. Correlation between variants, sample overlap, and weak first stages also affect estimation.

For a categorical instrument $Z$ indexing encouragement environments, this can be seen directly. Write $\Delta=Y^1-Y^0$, $\theta=\E(\Delta)$, and let $A(z)$ be treatment in environment $z$. Under joint exogeneity and exclusion, for two environments $s,r$ with a nonzero treatment contrast,
write $\psi_{s,r}(P)$ for the corresponding Wald ratio. Then
\begin{equation}\label{eq:ivdifference}
 \psi_{s,r}(P)=\theta+
 \frac{\Cov\{A(s)-A(r),\Delta\}}{\E\{A(s)-A(r)\}}.
\end{equation}
This identity follows by writing $Y=Y^0+A(Z)\Delta$. Equal effects are sufficient but unnecessary for equality with the average effect: it suffices that the compliance shift $A(s)-A(r)$ be uncorrelated with $\Delta$. Pairwise monotonicity instead gives pairwise complier interpretations. Neither interpretation should be inferred merely from differences in observed treatment prevalence across hospitals or studies; those differences may accompany changes in case mix or care.

Judge or examiner leniency fits this family when assignment is plausibly random within the appropriate strata, exclusions are defensible, and the relevant decision margin is monotone. Grouping judges differently can change both the instrument and the affected population. Estimated leniency scores also introduce estimation and dependence issues; changing their construction does not create new randomized assignments.
\structure Each instrument targets its own complier population. Under valid identification, the target difference from the population average is the covariance term in \eqref{eq:ivdifference}; disagreement among instruments is then expected rather than diagnostic. Instruments resting on the same exclusion argument share its failure. When the latent index model holds, the complier effects are points of one marginal treatment effect curve (Section~\ref{sec:average}).

\begin{example}[Shift-share instruments]\label{ex:bartik}
Consider the instrument $Z=\sum_{k=1}^K g_ks_k$, where $s_k$ are exposure shares and $g_k$ are fixed shocks for the algebra below. After any common linear residualization, define the share-specific ratio
\[
 \psi_k=\frac{\Cov(s_k,Y)}{\Cov(s_k,A)}
\]
whenever its denominator is nonzero. Then
\begin{equation}\label{eq:bartik}
\begin{aligned}
 \frac{\Cov(Z,Y)}{\Cov(Z,A)}&=\sum_k\alpha_k\psi_k,\\
 \alpha_k&=\frac{g_k\Cov(s_k,A)}{\sum_\ell g_\ell\Cov(s_\ell,A)}.
\end{aligned}
\end{equation}
When all included share ratios are defined and the pooled denominator is nonzero, $\sum_k\alpha_k=1$. This decomposition underlies the exposure design interpretation of \citet{gps2020}.
\end{example}
The weights can be negative, and exposure shares are not disjoint strata of individuals. Consequently \eqref{eq:bartik} is not a population partition or a generic collapsibility identity. A share with zero first stage but nonzero reduced form must be retained in the pooled numerator rather than assigned an undefined ratio. Identification through quasi-random shocks provides another justification for a shift-share analysis \citep{bhj2022}; its shock-level representation can be algebraically equivalent to the original estimator. The assignment and dependence structure of shocks also matters for inference, so an independent individual approximation is not generally appropriate.
\structure This is one estimator written in two ways, not a collection of routes: the share-level and shock-level representations are the same functional and should not be counted twice. The weights $\alpha_k$ are an implied measure in the sense of Section~\ref{sec:average}, and their possible negativity is what prevents reading the pooled ratio as an average effect.

\subsection{Thresholds and policy margins}
\begin{example}[Multiple cutoffs]\label{ex:rd}
Consider sites with different thresholds. Observe $O=(S,R,A,Y)$, where site $S=j$ assigns treatment at cutoff $c_j$ of running variable $R$. In a sharp design, $A=\ind(R\ge c_j)$ within site. If both potential outcome regression functions are continuous at $c_j$, with positive density of $R$ on both sides, then
\begin{equation}\label{eq:rd}
\begin{aligned}
 \psi_j(P)&=\lim_{r\downarrow c_j}\E(Y\mid R=r,S=j)\\
          &\quad-\lim_{r\uparrow c_j}\E(Y\mid R=r,S=j)\\
 &=\E(Y^1-Y^0\mid R=c_j,S=j).
\end{aligned}
\end{equation}
The target is understood through continuous versions at the cutoff. In a fuzzy design, divide the outcome jump by the nonzero treatment probability jump. Local exclusion, continuity of the relevant potential treatment and outcome distributions, and monotonicity then identify a local complier effect \citep{hahn2001identification}.
\end{example}
Multiple cutoffs create multiple local targets \citep{cattaneo2016multiple}. Their disagreement can reveal genuine differences across sites, baseline risks, or policy margins. Density continuity is a useful diagnostic implication in some assignment models, but it is not a substitute for continuity of potential outcomes. Likewise, a common numerical cutoff across sites does not make the local populations identical.

Changing a bandwidth around one cutoff typically changes an estimator of the same limiting jump. A fixed window estimand is another matter: it requires a local randomization or other explicit interpretation. Pooling bandwidth estimates as if they were independent identifying designs confuses tuning with identification. Regression kink designs similarly use changes in slopes, with derivative and assignment restrictions replacing the jump conditions; the local response they identify must be defined before comparing different kinks. Bunching designs require their own behavioral and counterfactual density restrictions and cannot be added to an RD catalog solely because both involve a policy threshold.
\structure Each site and cutoff has its own target; differences among them are target differences under valid continuity. A pooled estimate across cutoffs carries the density-based measure of Section~\ref{sec:average}; a policy summary requires a measure over cutoffs chosen for the policy question.

\subsection{Cohorts and eligibility occasions}
\begin{example}[Staggered adoption]\label{ex:staggered}
Observe a panel $(G,Y_1,\ldots,Y_T)$, where $G=g$ denotes first treatment at time $g$ and $G=\infty$ denotes no treatment. Let $Y_t(g)$ be the outcome under adoption at $g$. For $t\ge g$, no anticipation and parallel untreated trends between cohort $g$ and never-treated units imply
\begin{equation}\label{eq:staggered}
\begin{aligned}
 \psi_{g,t}(P)&=\E(Y_t-Y_{g-1}\mid G=g)\\
             &\quad-\E(Y_t-Y_{g-1}\mid G=\infty)\\
 &=\E\{Y_t(g)-Y_t(\infty)\mid G=g\}.
\end{aligned}
\end{equation}
This identifies $\theta_{g,t}$, the group time effect of \citet{callaway2021difference}, in its unconditional form with a pretreatment baseline and available never-treated controls.
\end{example}
Two axes are changing: who is treated and how long treatment has operated. Equality across $(g,t)$ is a substantive homogeneity restriction. By contrast, alternative valid control groups for a \emph{fixed} $(g,t)$, such as never-treated and not-yet-treated units, provide alternative identification of that cell's effect. Keeping these two indices conceptually separate avoids treating event time dynamics as evidence of design failure.
\structure The collection combines target heterogeneity across $(g,t)$ with identification multiplicity within each cell. A summary is a projection: a policy measure $q_{g,t}$ gives $\sum_{g,t}q_{g,t}\theta_{g,t}$, and the event time family $m_\beta(g,t)=\beta_{t-g}$ with cohort size weights gives event time summaries of the kind used by \citet{callaway2021difference}. The measure should specify the policy population and follow-up distribution; it should not be inherited silently from a two-way fixed effects coefficient.

\begin{example}[Sequential target trials]\label{ex:trials}
Suppose longitudinal data permit a trial at each occasion $t$. Let $I_t$ indicate eligibility, $H_t$ denote history before assignment, and $A_t$ denote the treatment decision. Let $Y_t^a$ be the outcome at a fixed horizon after that occasion had the decision at $t$ been set to $a$, with later treatment then following its natural course; this is the treatment policy contrast of initiating versus not initiating at $t$. With consistency, exchangeability given $(H_t,I_t=1)$, and positivity,
\[
\begin{aligned}
 \psi_t(P)&=\E\{\mu_{1t}(H_t)-\mu_{0t}(H_t)\mid I_t=1\}\\
 &=\E(Y_t^1-Y_t^0\mid I_t=1)=\theta_t,
\end{aligned}
\]
where $\mu_{at}(h)=\E(Y_t\mid A_t=a,H_t=h,I_t=1)$ and $Y_t$ denotes the corresponding observed follow-up outcome. A contrast that also fixes later treatment, such as sustained initiation, is a different intervention: it requires the longitudinal versions of these conditions and a longitudinal identification formula.
\end{example}
Eligibility can change with disease course, previous treatment, and survival. Thus these trials need not identify one effect. Uniform weights over trial occasions and weights proportional to eligibility probabilities define, respectively,
\begin{equation}\label{eq:eligibility}
 \theta_{\mathrm{uniform}}=T^{-1}\sum_t\theta_t,
 \qquad
 \theta_{\mathrm{eligible}}=
 \frac{\sum_t P(I_t=1)\theta_t}{\sum_t P(I_t=1)}.
\end{equation}
\citet{gervasoni2026} develop this estimand first perspective; they also standardize each occasion's conditional effects to the baseline covariate distribution, which defines a third target when baseline is meaningful. The second quantity refers to a distribution of eligible person-occasions. A person eligible ten times contributes differently from one eligible once. It is not automatically a person-level average effect. If the design instead assigns each person to their first eligible occasion, the groups form a partition; aggregation can then recover the effect of a policy defined relative to that occasion.

\structure Each occasion targets the effect among those eligible then; differences across occasions are target differences. The summaries in \eqref{eq:eligibility} are constant projections under two measures, whereas a pooled regression imposes a variance-based measure (Section~\ref{sec:average}). Repeated participation makes the person, not the person-occasion, the independent unit.

\subsection{Principal strata and outcomes truncated by death}
\label{sec:principal}
Principal stratification \citep{frangakis2002} provides a useful boundary case, and it resolves the fifth analysis of the introduction. Let $D^a$ indicate survival to the outcome visit under treatment $a$; the outcome $Y^a$ is then defined only when $D^a=1$. In a randomized comparison, observed treated survivors represent the stratum $\{D^1=1\}$ and observed untreated survivors the stratum $\{D^0=1\}$. These differ whenever treatment affects survival, so the observed contrast among survivors compares two populations and need not be a causal effect in either. A causal effect is defined in the principal stratum of those who would survive under both treatments,
\[
 \theta_{\mathrm{SACE}}=\E(Y^1-Y^0\mid D^1=D^0=1).
\]
That stratum is latent. Randomization identifies $P(D^a=1)$ for each $a$ but not the joint distribution of $(D^0,D^1)$, so $\theta_{\mathrm{SACE}}$ is not identified without further assumptions; \citet{zhangrubin2003} derive large sample bounds for effects in principal strata, with or without additional assumptions such as monotonicity of survival. Alternatively, the question can be redefined through an outcome that incorporates death. That yields a different target, not an identified version of the same one. The survivor analysis is therefore neither an alternative route to the population effect nor a local effect of the kind identified by instruments or cutoffs: its causal target is undefined until the question is refined.

More generally, strata defined by joint potential intermediate outcomes can partition a population, and their effects can have scientific meaning. But the partition is generally latent, and randomization alone need not identify every stratum effect. Listing principal strata therefore does not manufacture multiple identifying functionals. Identification and the intended weighted target must be established separately.

\section{Aligning sources and outcomes}
\label{sec:sources}
A common target is sometimes obtained by standardizing every source to one population. Here the outer distribution is part of the question, while the conditional outcome information can come from different places. The sources are sampled separately, so the sampling model of Section~\ref{sec:language} applies.

\begin{example}[Several trials or observational sources]\label{ex:transport}
Observe data indexed by source $S$, with baseline covariates $X$ available in a target population whose distribution is $F_\star$. In source $j$, suppose conditional treatment exchangeability given $X$, consistency, and positivity identify
$\delta_j(x)=\E(Y\mid A=1,X=x,S=j)-\E(Y\mid A=0,X=x,S=j)$.
If conditional treatment effects transport,
$\E(Y^1-Y^0\mid X=x,\star)=\E(Y^1-Y^0\mid X=x,S=j)$,
and target covariate support is covered by source $j$, then
\begin{equation}\label{eq:transport}
 \psi_j(P)=\int\delta_j(x)\,dF_\star(x)=\theta_\star.
\end{equation}
The observed data include the source samples and the target covariate sample, or $F_\star$ can be specified externally.
\end{example}
Conversely, standardization does not remove a difference in treatment versions or outcome ascertainment. The choice of source mixture in the collected dataset is not necessarily a population distribution of scientific interest. \citet{ung2026} emphasize making the target population and sampling model explicit when combining information; reviews of transportability and of combining trials with observational data develop these conditions in detail \citep{pearlbareinboim2014,colnet2024,degtiar2023}.
\structure After standardization every source targets $\theta_\star$. Unstandardized source effects $\int\delta_j\,dF_j$, with $F_j$ the covariate distribution of source $j$, differ from $\theta_\star$ by $\int\delta_j\,d(F_j-F_\star)$ when transport holds, so their disagreement can be a pure target difference. The discrepancy of the standardized analysis is the failure of transport for source $j$; sources recruited through the same mechanism can share it.

\begin{example}[Several external control sources]\label{ex:external}
Let $S=0$ denote an index randomized trial and $S=j>0$ an external source providing untreated outcomes. Write
$m_{10}(x)=\E(Y\mid A=1,X=x,S=0)$ and
$m_{0j}(x)=\E(Y\mid A=0,X=x,S=j)$.
The index trial average effect is identified by
\begin{equation}\label{eq:external}
 \psi_j(P)=\int\{m_{10}(x)-m_{0j}(x)\}\,dP(x\mid S=0)
\end{equation}
if the index treatment arm identifies the treated mean, source $j$ identifies its untreated mean, and that untreated mean transports to the trial population. These conditions include consistency across sources, untreated exchangeability in a nonrandomized external source, and the necessary source and treatment overlap. The trial's own control arm supplies an internal version of the same contrast.
\end{example}
Only the untreated mean must transport in this example, but that is still a consequential assumption. Blinding, background care, recruitment, calendar time, and measurement procedures can change it. \citet{senn2026external} argues that these study effects must be confronted explicitly when borrowing controls. An external source that matches patient-level covariates can still differ in the care delivered to those patients. Several sources sharing the same historical practice may agree for the same wrong reason.

This is also a setting in which an anchor changes the interpretation of disagreement. When the internal randomized comparison is credible, discrepancies from external control analyses are informative relative to that anchor. A symmetric pooling rule that treats all sources as equally uncertain can discard an asymmetry supplied by the design. The anchor itself does not answer every question: its population may differ from the intended clinical population, and follow-up or adherence may require further assumptions.
\structure All analyses target the index trial effect. The discrepancy of source $j$ is $\int\{\E(Y^0\mid X=x,S=0)-m_{0j}(x)\}\,dP(x\mid S=0)$, and the internal control arm has zero discrepancy by randomization. Every analysis shares the trial's treated arm.

\begin{example}[Several surrogate sets]\label{ex:surrogate}
Let $S=E$ denote a trial with $(X,A,M)$ observed and $S=V$ a validation source with $(X,M,Y)$ observed. The goal is the trial population effect on the long-term outcome $Y$. For a surrogate set $M_j$, define $h_j(m,x)=\E(Y\mid M_j=m,X=x,S=V)$. Under trial treatment exchangeability, mean surrogacy in the trial,
\[
 \E(Y\mid A,M_j,X,S=E)=\E(Y\mid M_j,X,S=E),
\]
and comparability of the conditional outcome mean across the two sources, the long-term average effect is identified by
\begin{equation}\label{eq:surrogate}
 \psi_j(P)=\E\{\eta_{1j}(X)-\eta_{0j}(X)\mid S=E\},
\end{equation}
where $\eta_{aj}(x)=\E\{h_j(M_j,x)\mid A=a,X=x,S=E\}$.
Consistency and support overlap for all transported evaluations are also required \citep{athey2024surrogate}.
\end{example}
Different surrogate sets or validation sources can yield different functionals for the same long-term effect. A short-term biomarker effect is not itself the long-term clinical effect: it becomes informative about that effect through the surrogate assumptions. If direct long-term outcomes later become available in the trial, they can provide an anchor under a different assumption set.
\structure All surrogate sets target the same long-term effect. The discrepancy of set $j$ comes from failures of surrogacy or comparability; a pathway from treatment to outcome missed by every surrogate set is a common discrepancy that agreement among surrogate analyses cannot reveal.

\section{Multiplicity in clinical evidence}
\label{sec:clinical}
Clinical discussions of multiplicity often concern several endpoints, doses, time points, subgroup claims, or interim looks. These may involve several causal questions. By contrast, the adjustment sets in Example~\ref{ex:adjustment} can address one question through several identifying assumptions. The distinction does not exempt either situation from accounting for selection. It tells us what must be calibrated.

Suppose $p_1,\ldots,p_J$ are valid tests of the same null under a shared maintained model. Rejecting if any $p_j\le\alpha$ uses an opportunity to reject $J$ times; the union bound gives an error probability no greater than $J\alpha$, and a Bonferroni threshold $\alpha/J$ controls it at $\alpha$ without independence. This calculation says nothing about unmeasured confounding. If a test is invalid under the actual identifying assumptions, dividing its threshold by $J$ does not restore a causal interpretation.

There is a different rule when robustness to identification assumptions is intended. If under the causal null at least one prespecified test is valid, rejecting only when \emph{every} test rejects at level $\alpha$ has error probability at most $\alpha$: rejection implies rejection by the valid test. This is the intersection--union principle \citep{berger1982,bergerhsu1996}, applied to a null that is the union over strategies of ``no effect and strategy $j$ is valid.'' It can have low power. Evidence factors \citep{rosenbaum2011,karmakar2019} and tests that combine several candidate causal models \citep{yang2024robust} pursue related guarantees under conditions on the joint behavior of the component tests. The guarantee depends on the tests addressing the same causal null under their respective models. A test of an instrumental variable complier effect and a test of a population average effect do not generally meet that requirement. Claims about a single common effect, a global absence of any effect, and a sharp null for every individual must be distinguished.

A third task is discovery. Searching trial data for treatment effect modifiers can generate useful hypotheses for later investigation. As \citet{hahn2026} argues in an essay, this purpose differs from demonstrating a primary effect. The discovery should be reported as such, with attention to the selection procedure and subsequent validation. A significant effect in one subgroup and a nonsignificant effect in another is not itself evidence of an interaction; the difference in effects is the relevant contrast. Nor does heterogeneity in conditional mean effects identify the distribution of individual treatment effects.

Selection also enters through design decisions. Investigators may choose a covariate set, revise eligibility, select a control source, or retain the specification with the smallest standard error. \citet{stensrud2026} discusses separating protocol specification from emulation by sample splitting. The principle applies more widely: an inference method for a prespecified collection does not automatically cover a collection chosen after examining its results. Cross-fitting nuisance estimators is not, by itself, a remedy for selecting the scientific target or strategy catalog on the same data. Multiverse and specification curve displays \citep{steegen2016,simonsohn2020} make such choices visible; the account of Section~\ref{sec:collection} adds which of the displayed specifications share a target and which share an assumption.

For practice, a useful report begins with the intervention, population, outcome, follow-up, and effect scale. It records each candidate route and its assumptions before presenting estimates. Analyses that change the target are labeled accordingly; algebraically equivalent formulas are grouped together. The joint sampling covariance accompanies comparisons when it can be estimated. Finally, the report states what is being done with the collection: a diagnostic comparison, an efficiency combination, a declared projection, a sensitivity analysis, a confirmatory decision, or exploration for a new study. These purposes can coexist, but their guarantees should not be exchanged.

For the motivating treatment study, this would place the two adjustment analyses in one group. The comparator analysis targets the treated population; it joins that group only if the adjustment analyses are standardized to the treated covariate distribution, so that all three address the effect on the treated. Physician preference analyses would retain their complier interpretation unless additional assumptions aligned it with the main population. The survivor comparison would be set apart: it becomes a causal analysis only after the question is redefined, either as an effect among those who would survive under both treatments, with bounds or additional assumptions, or through an outcome that incorporates death. A summary would then explain which assumptions the comparisons challenge and which remain shared. The aim is to make the next scientific decision easier: refine measurement, improve overlap, obtain an experimental anchor, or design a new study to distinguish rival explanations.

\section{Discussion}
\label{sec:discussion}
Identification multiplicity is a property of the connection between a question, assumptions, and data. It cannot be read off the number of estimators in an analysis script. Across fourteen designs, the same three structures recur: a map from analyses to targets, an overlap in assumptions, and a dependence among estimates. Alternative covariate sets, comparators, proxy roles, and mediator sets can identify one target; instruments, thresholds, cohorts, and eligibility occasions commonly identify a curve of targets; and source combination requires an explicit account of transport and sampling.

The operations available on a collection match these structures. Contrasts address differential discrepancies and are blind to a common one. Covariance-aware combination improves efficiency only under a common valid target. Aggregation has a causal meaning when it is a declared projection of an identified causal curve, and not otherwise. Unions of intervals, bias bounds, and sign restrictions represent weaker premises about validity. Each operation is a tool for a specific question, and its guarantee does not transfer to the others.

This organization also identifies useful directions for methodological work. Strategy collections should be described by their shared assumptions and target populations, not only by a covariance matrix. Procedures should distinguish a credible anchor from a symmetric set of uncertain routes. Their behavior should be stable under duplication or harmless re-expression of an analysis. When the number of strategies grows, dependence and estimation error must be controlled at the level where information actually enters: people, clusters, environments, or shocks. Sensitivity analysis should address common bias as well as differences among strategies, and the measures implied by standard pooled estimators deserve to be reported as routinely as their standard errors.

Several routes to a causal answer are most useful when their differences are understood. Agreement is then interpreted in light of shared weaknesses; disagreement can distinguish target heterogeneity from conflicting assumptions; and an aggregate is tied to a population or decision that gives its weights meaning. The result is a more informative account of the evidence, including what the available designs leave unresolved.

\bibliographystyle{spr-ims-nameyear}
\bibliography{references}
\end{document}